# Non-Fourier heat transport in metal-dielectric core-shell nanoparticles under ultrafast laser pulse excitation

M. Rashidi-Huyeh[a, b, c], S. Volz[a] and B. Palpant[b]

[a] Laboratoire d'Energétique Moléculaire et Macroscopique, Combustion (EM2C), CNRS – Ecole Centrale de Paris, Grande Voie des Vignes, 92295 Châtenay-Malabry, France

[b] Université Pierre et Marie Curie – Paris 6, UMR 7588, INSP, Campus Boucicaut, 140 rue de Lourmel, Paris, F-75015 France

[c] Department of Physics, University of Sistan and Baluchistan, Zahedan, Iran

## *Abstract*

Relaxation dynamics of embedded metal nanoparticles after ultrafast laser pulse excitation is driven by thermal phenomena of different origins the accurate description of which is crucial for interpreting experimental results: hot electron gas generation, electron-phonon coupling, heat transfer to the particle environment and heat propagation in the latter. Regardingthis last mechanism, it is well known that heat transport in nanoscale structures and/or at ultrashort timescales may deviate from the predictions of the Fourier law. In these cases heat transport may rather be described by the Boltzmann transport equation.

We present a numerical model allowing us to determine the electron and lattice temperature dynamics in a spherical gold nanoparticle core under subpicosecond pulsed excitation, as well as that of the surrounding shell dielectric medium. For this, we have used the electron-phonon coupling equation in the particle with a source term linked with the laser pulse absorption, and the ballistic-diffusive equations for heat conduction in the host medium. Either thermalizing or adiabatic boundary conditions have been considered at the shell external surface.

Our results show that the heat transfer rate from the particle to the matrix can be significantly smaller than the prediction of Fourier's law. Consequently, the particle temperature rise is larger and its cooling dynamics might be slower than that obtained by using Fourier's law. This difference is attributed to the nonlocal and nonequilibrium heat conduction in the vicinity of the core nanoparticle. These results are expected to be of great importance for analyzing pump-probe experiments performed on single nanoparticles or nanocomposite media.

## *I.  Introduction*

Noble metal nanoparticles are especially interesting because of their linear and nonlinear optical properties linked with the surface plasmon resonance.[1–3] Thanks to these properties, numerous applications are being developed based on metal nanoparticles, such as photonic devices, molecular sensing, biological cell imaging or photothermal therapy.[4–6] These specific properties are driven by the coupling between light, electrons and phonons within particles as well as by energy exchanges between particles and their environment, the study of which has motivated numerous fundamental investigations for about two decades. The optical properties of nanocomposite materials consisting of noble metal nanoparticles embedded in a dielectric matrix can be modified by thermal phenomena of different nature – such as electron-electron and electron-phonon interaction, heat transfer to the surrounding medium, and heat exchange between neighboring nanoparticles – when exposed to laser light.[7–14] The optimal experimental tool for studying these mechanisms is certainly pump-probe time-resolved spectroscopy using ultrashort laser pulses.[13] In such experiments a strong laser pulse induces a series of coupled optical and thermal phenomena, leading to the transient modification of the material optical response. The latter mainly depends on the energy distribution of the conduction electron gas, the dynamics behavior of which is particularly sensitive to energy exchange between the particle and its environment as well as the energy transport in the latter. In this case, Fourier's law is no longer suited to describe heat propagation as the spatial and time scales under consideration are smaller than the heat carrier mean free path and lifetime, respectively.[16–19]

Beyond the interest raised by metal nanoparticle optical properties themselves, it is noteworthy that inclusion of such particles in a transparent host has been proposed as a model method to study the impact of the presence of defects in transparent media on their damage threshold when using high-intensity pulsed-laser radiation.[20,21] The analysis of the energy exchanges in such systems takes then also on particular importance for technological purposes.

The two-temperature model was previously applied for bulk metal and thin films[11,14] to model the electron relaxation dynamics related to the observed optical response. This model only consists in describing the electron-lattice coupling in the metallic nanoparticle, the heat transferred to the matrix being neglected. In a previous work,[7] we have presented an improvement of this approach to determine the temperature dynamics in nanocomposite materials under pulsed laser (three-temperature model). In this model, the electron excitation by the light pulse, the usual electron-phonon coupling in metal particles, the particle-matrix thermal transfer at the interface and the heat diffusion predicted by Fourier's law in the matrix were considered together. However, as emphasized above, in the general case the heat transfer cannot be correctly described by the Fourier approach when the medium characteristic spatial scale (and/or the characteristic time of the heat variation) becomes as small as the heat carrier mean free path (and/or the heat carrier lifetime). In these cases, the heat transfer in the medium might rather be described by the Boltzmann transport equation (BTE).[12] Chen has proposed the use of an alternative equation, named the ballistic-diffusive equation (BDE), which is based on the BTE within the relaxation time approximation and appears more simple to solve.[17,18] In this approach the heat flux intensity and the internal energy at any point of the matrix are splitted into two components, corresponding to ballistic carriers originating from the boundaries and to scattered carriers, respectively.[16–19]

In this paper we apply our previous three-temperature model to a gold spherical nanoparticle core surrounded by an alumina shell by using the BDE rather than the Fourier law to describe the particle-matrix energy exchange and heat transport in the matrix. The dynamics of both the metal electron and lattice temperatures, as well as the internal energy at each point of the host medium, will be determined. Note that, due to the ambiguity inherent to the definition of a temperature at a spatial scale lower than the heat carrier mean free path in metal nanoparticles, the lattice temperature will rather be considered as a measure of the internal energy. Both thermalizing (imposed temperature at the shell outer surface) and adiabatic (isolated core-shell) boundary conditions will be examined. The influence of the nanometric scale on the heat transport mechanisms will be highlighted by comparing, for different shell thicknesses, the results with those obtained with the classical Fourier law.

## II. Model

The core-shell nanoparticle consists of a metal nanosphere with radius $R_p$ surrounded by a dielectric shell with internal and external radiuses $R_p$ and $R_{ex}$, respectively (Fig. 1). The shell thickness is then $d_s = R_{ex} - R_p$. Such core-shell nanoparticles have already been synthesized[22] and investigated for their linear and nonlinear optical properties.[23,24] The case of an isolated nanoparticle embedded in an host medium, or the one of a diluted nanocomposite material, corresponds to the infinite shell thickness limit and can also be accounted for by the present model. As we have already described in Ref. 7, a light pulse is partly absorbed by the conduction electron gas of the metal particle. The room-temperature Fermi-Dirac electron energy distribution is then changed into a low-density nonthermalized highly energetic electron population. The electron-electron collisions within the electron gas then leads to the formation of a hot Fermi-Dirac distribution.[11,12] This internal thermalization occurs on a subpicosecond time scale depending on the particle size as well as on the energy amount absorbed. Note that we have recently developed a theoretical method to account for this nonthermal regime, which could be integrated into the present model. However, for the sake of simplification, we will disregard this refinement here by considering an instantaneous internal energy redistribution within the electron gas. An electron temperature, $T_e$, can then be defined at every time $t$. The electron energy is then transferred to the particle lattice via electron-phonon coupling and is then released into the dielectric shell. We shall consider the same assumptions as in reference 7. The evolution of the electron temperature is then driven by the following equation:

$$C_e \frac{\partial T_e}{\partial t} = -G(T_e - T_l) + P_{vol}(t), \quad (1)$$

where $C_e = \gamma_e T_e$ is the electron heat capacity, $\gamma_e$ is a constant, $G$ is the electron-phonon coupling constant, $P_{vol}(t)$ denotes the instantaneous power absorbed per metal volume unit and $T_l$ is the lattice temperature. The temporal evolution of $T_l$ is controlled, on one hand, by the energy received from the electron gas via electron-phonon coupling, and on the other hand by the heat transferred to the shell medium at the interface. We therefore write:

$$V_p C_l \frac{\partial T_l}{\partial t} = V_p G(T_e - T_l) - H(t), \quad (2)$$

where $C_l$ is the heat capacity of the lattice, $V_p$ is the volume of the particle and $H(t)$ is the instantaneous heat power transferred to the dielectric shell from the particle. $H(t)$ may be written as a function of the heat flux density, $\mathbf{q}(\mathbf{r},t)$, integrated over the particle surface, $S_p$:

$$H(t) = \int_{S_p} \mathbf{q}(\mathbf{r},t) \cdot \mathbf{n} \, ds , \qquad (3)$$

where $\mathbf{n}$ is the outward unit vector normal to the particle surface. In the aim to determine $\mathbf{q}(\mathbf{r},t)$, we have applied the ballistic-diffusive approximation[18] that will be explained in the following section.

### a. Ballistic-Diffusive equations

The essence of the ballistic-diffusive approximation stands on the splitting of the carrier heat distribution function at any point $\mathbf{r}$ of the matrix, $f(t,\mathbf{r},\mathbf{\Omega})$, in a given direction $\mathbf{\Omega}$, into two components: One is due to carriers emitted directly from the boundaries, without any scattering, and represents the ballistic component. The other is related to the carriers which arrive after scattering (or emission after absorption) from other points of the medium. Their behavior may be described by a diffusive process. This approach is based on the Boltzmann transport equation (BTE) under the relaxation time approximation[15] and can be expressed by the phonon intensity at point $\mathbf{r}$ and direction $\mathbf{\Omega}$ defined as:

$$I_\omega(t,\mathbf{r},\mathbf{\Omega}) = |\mathbf{v}| \hbar \omega \, f(t,\mathbf{r},\mathbf{\Omega}) \rho(\omega)/4\pi \qquad (4)$$

where $\mathbf{v}$ is the phonon group velocity, $\omega$ is the phonon circular frequency, and $\rho(\omega)$ is the phonon density of states per unit volume. The instantaneous heat flux at each point, $\mathbf{q}(\mathbf{r},t)$, may be obtained by integration of $I_\omega(t,\mathbf{r},\mathbf{\Omega})$ over all phonon frequencies and all directions (solid angle). The ballistic flux component can then be written as:

$$\mathbf{q_b}(\mathbf{r},t) = \int_0^\infty \left[ \int I_{b\omega}(t,\mathbf{r},\mathbf{\Omega}) \mathbf{\Omega} \, d\mathbf{\Omega} \right] d\omega \qquad (5)$$

where $I_{b\omega}(t,\mathbf{r},\mathbf{\Omega})$ is the ballistic intensity at $\mathbf{r}$ in direction $\mathbf{\Omega}$ due to the boundary intensity, $I_{w\omega}$ ($w$ stands for "wall"), at $\mathbf{r} - r'\mathbf{\Omega}$:

$$I_{b\omega}(t,\mathbf{r},\mathbf{\Omega}) = I_{w\omega}\left(t - \frac{r'}{|\mathbf{v}|}, \mathbf{r} - r'\mathbf{\Omega}, \mathbf{\Omega}\right) \exp(-r'/\Lambda_\omega). \qquad (6)$$

$r'$ is the distance between the boundary point corresponding to the direction $\mathbf{\Omega}$ and $\mathbf{r}$ (see Fig. 1) and $t - r'/|\mathbf{v}|$ represents the time retardation related to the finite speed of phonons. The equation governing the internal ballistic energy component, $u_b$, may be given by:

$$\tau \frac{\partial u_b}{\partial t} + \nabla \cdot \mathbf{q_b} = -u_b \tag{7}$$

where $\tau$ is an average of the phonon relaxation time in the matrix.

The diffusive part is an approximation modeling the heat carriers undergoing many collisions before reaching the volume element. It may be described by the following equation:

$$\tau \frac{\partial \mathbf{q_m}}{\partial t} + \mathbf{q_m} = -\frac{\kappa_d}{C_d} \nabla u_m \tag{8}$$

where $\mathbf{q_m}$ and $u_m$ represent the diffusive components of heat flux and internal energy, respectively, $\kappa_d$ is the heat conductivity of the medium and $C_d$ denotes its heat capacity. At last, the energy conservation relation imposes:

$$\frac{\partial u}{\partial t} + \nabla \cdot \mathbf{q} = 0, \tag{9}$$

$u = u_b + u_m$ and $\mathbf{q} = \mathbf{q_b} + \mathbf{q_m}$ being the total internal energy and heat flux, respectively. The following hyperbolic equation for $u_m$ is finally reached:

$$\tau \frac{\partial^2 u_m}{\partial t^2} + \frac{\partial u_m}{\partial t} = \nabla \cdot \left( \frac{\kappa_d}{C_d} \nabla u_m \right) - \nabla \cdot \mathbf{q_b}. \tag{10}$$

The boundary conditions ensure that heat carriers originating from the boundaries have ballistic components only. This leads to the boundary condition for the diffusive component:[18]

$$\tau \frac{\partial u_m}{\partial t} + u_m = -\frac{2\Lambda_d}{3} \nabla u_m \cdot \mathbf{n} \tag{11}$$

where $\mathbf{n}$ is the inward unit vector perpendicular to the boundary and $\Lambda_d = |\mathbf{v}|\tau$ is the mean free path of heat carriers in the dielectric. $\Lambda_d$ may be given by $\sqrt{3 D_d \tau}$ with $D_d$ being the heat diffusivity of the dielectric medium.

## b. Application to a core-shell nanoparticle

We consider the shell initially at ambient temperature $T_0$. At time $t=0$, the inner shell surface, *i.e.* at $r=R_p$, emits phonons at core lattice temperature, $T_l(t)$. As such a core-shell nanoparticle presents a spherical symmetry, the heat flux as well as the local internal energy (or local temperature) only depend on the radial distance and the heat flux is a radial vector. Applying equations (5) and (6), the ballistic component of the heat flux $q_b(t,\mathbf{r})$ at a given point *M* of the matrix (Fig. 1), with a position vector **r**, at time *t*, is provided by:

$$q_b(t,\mathbf{r}) = \frac{C_d |\mathbf{v}|}{2} \int_{\alpha_1(r)}^{1} \Delta T_l(t - r'/|\mathbf{v}|) \exp(-r'/\Lambda_d) \alpha_p d\alpha_p + q_{0b}(\mathbf{r}). \quad (12)$$

$r'$ is the distance between a point $s'$ of the particle surface and *M*, $\alpha_p = \cos(\theta_p)$, where $\theta_p$ is the polar angle of $s'$ relative to (*MO*) (Fig. 1), and $\Delta T_l(t) = T_l(t) - T_0$ is the lattice temperature rise. $q_{0b}(\mathbf{r})$ is a time independent term which is given by:

$$q_{0b}(\mathbf{r}) = \frac{C_d |\mathbf{v}| T_0}{2} \left[ \int_{\alpha_1(r)}^{1} \exp(-r'/\Lambda_d) \alpha_p d\alpha_p + \int_{\alpha_2(r)}^{1} \exp(-r''/\Lambda_d) \alpha_{ex} d\alpha_{ex} \right] \quad (13)$$

where $r''$ is the distance of a point $s''$ of the external boundary surface and $\alpha_{ex} = \cos(\theta_{ex})$, $\theta_{ex}$ being the polar angle defined from (*MO*) as illustrated in Fig. 1. The integrations in the two last equations cover all the points of the boundary surfaces (particle and shell external surface) which can be viewed from *M*.

Following the same procedure, the instantaneous ballistic component of internal energy at point **r** may be given as:

$$u_b(t,\mathbf{r}) = \frac{C_d |\mathbf{v}|}{2} \int_{\alpha_1(r)}^{1} \Delta T_l(t - r'/|\mathbf{v}|) \exp(-r'/\Lambda_d) d\alpha_p + u_{0b}(\mathbf{r}), \quad (14)$$

with:

$$u_{0b}(\mathbf{r}) = \frac{C_d |\mathbf{v}| T_0}{2} \left[ \int_{\alpha_1(r)}^{1} \exp(-r'/\Lambda_d) d\alpha_p + \int_{\alpha_2(r)}^{1} \exp(-r''/\Lambda_d) d\alpha_{ex} \right] \quad (15)$$

Finally Eq. (8) can be rewritten as:

$$\tau \frac{\partial q_m}{\partial t} + q_m = -\frac{\kappa_d}{C_d} \frac{\partial u_m}{\partial r}. \quad (16)$$

190   **Initial conditions.** As the system is initially at ambient temperature, $T_0$, the initial conditions for the shell medium are:

$$t = 0: \quad T(r,0) = T_0, \quad \left.\frac{\partial T(r,t)}{\partial t}\right|_{t=0} = 0, \tag{17}$$

or in term of internal energy:

$$t = 0: \quad u(r,0) = u_m(r,0) + u_b(r,0) = C_d T_0, \quad \left.\frac{\partial u(r,t)}{\partial t}\right|_{t=0} = 0, \tag{18}$$

195   **Boundary conditions.** Applying Eq. (11), the boundary condition equation for the diffusive component at the core particle surface may be given by:

$$r = R_p: \quad \tau \frac{\partial u_m}{\partial t} + q_m = \frac{2\Lambda_d}{3} \frac{\partial u_m}{\partial r}. \tag{19}$$

We have considered two kinds of boundary conditions at the outer shell surface ($r = R_{ex}$). The first one corresponds to a full thermalization condition. In this case the temperature at the
200   surface is imposed as the ambient temperature. Using Eq. (11), one can obtain the equation for the diffusive component at this surface:

$$r = R_{ex}: \quad \tau \frac{\partial u_m}{\partial t} + q_m = -\frac{2\Lambda_d}{3} \frac{\partial u_m}{\partial r}. \tag{20}$$

The second boundary condition considered here is adiabatic. In this case, the total heat flux vanishes at the surface. We have then:

205   $$\mathbf{q_m}(r = R_{ex}, t) = -\mathbf{q_b}(r = R_{ex}, t). \tag{21}$$

In order to determine the electron and lattice temperatures of the particle as well as the heat flux and local temperature in the shell, equations (1), (2), (12)–(16) and the energy conservation relation Eq. (9) associated to the initial and boundary conditions, presented above, should be solved.

210   ### III. Results and discussion

Results will be presented in terms of non-dimensional parameters as defined in the Appendix. The problem was solved for a gold nanoparticle core and an alumina (amorphous $Al_2O_3$) shell with thermodynamic properties corresponding to their bulk phase[7]: $G = 3 \times 10^{16} \text{ W m}^{-3} \text{ K}^{-1}$, $\gamma = 66 \text{ J m}^{-3} \text{ K}^{-2}$, $C_l = 2.49 \times 10^6 \text{ J m}^{-3} \text{ K}^{-1}$, $D_d = 1.16 \times 10^{-5} \text{ m}^2 \text{ s}^{-1}$, 215    $\kappa_m = 36 \, \text{m}^2 \, \text{s}^{-1}$ and $|\mathbf{v}| = 6400 \, \text{ms}^{-1}$. The free path of heat carriers, $\Lambda_d$, and phonon relaxation time, $\tau$, in the dielectric are then 5.4 nm and 0.85 ps, respectively. The initial temperature, $T_0$, is set to 300 K. Moreover, $P_{vol}(t)$ is considered to exhibit the same time dependence as the incident pulse, which is supposed to be a Gaussian. It is also to mention that as the spatial width of the pulse (at least few tens micrometers) is much larger than the particle size (from a

220    few nanometers to tenths of nanometers), the instantaneous energy absorbed by the particle is homogeneous over the particle volume. The expression of the absorbed power is as $P_{vol}(t) = A e^{-B(t-t_0)^2}$ where the parameter values are chosen equal to those of our earlier work[7] (*i.e.* $A = 1.4 \times 10^{21} \, \text{W m}^{-3}$, $B = 2.3 \times 10^{26} \, \text{s}^{-2}$ and $t_0 = 150 \, \text{fs}$). The corresponding pulse duration is 110 fs. The numerical method is based on the finite differential element technique

225    and the stability has been ensured by choosing the relation between $\delta r$ and $\delta t$ [*i.e.* $\delta t = (\delta r)^2 / (2 D_d)$][25] as provided by the diffusion mechanism. The calculation accuracy has been tested: By dividing the value of $\delta r$ by a factor two and $\delta t$ by a factor four, variations in the relative electron and lattice temperatures are lower than 2.5%.

### a. Thermalizing boundary condition

230    Let us first compare the results of the BDE model to the one based on Fourier's law. We have calculated the temperature dynamics using these two approaches with the same parameters as well as the same initial and boundary conditions. Figure 2 presents the electron and lattice nondimensional temperature dynamics (black and grey lines) obtained using BDE (solid line) and Fourier's law (dash line). The gold particle radius is 10 nm and the shell

235    thickness, $d_s$, is of the order of $\Lambda_d$ (~5.4 nm). It clearly appears that the two methods predict the same behaviour for the electron temperature during a few initial picoseconds, where $T_e$ increases very rapidly up to 2200 K ($\Theta_e = 6.3$) just after the pulse. It then reveals a rapid relaxation due to electron-phonon scattering. $T_l$ is very low as compared to $T_e$ during this time and can then be neglected in Eq. (1). On the other hand the pulse is quickly off at the

240    beginning of this short time regime. Consequently, one can easily show that the electron temperature may simply be obtained by:

$$T_e(t) - T_{e,\max} \simeq -\frac{G}{\gamma} t \quad (22)$$

where $T_{e,\max}$ is the maximum value of the electron gas temperature. The characteristic time of the rapid relaxation $\tau_r$, defined as the time for the electron temperature to reach the half of its maximum, is $\tau_r = \dfrac{T_{e,\max}\gamma}{2G}$. For the values considered here, this time is about ~ 4 ps. At the end of this relaxation, electrons and lattice attain a thermal equilibrium and we can indifferently denote their respective temperatures by the term *particle temperature*. Whatever the theoretical approach, this thermal equilibrium occurs after ~ 11 ps.

This rapid relaxation is then followed by a slow one linked with the heat transfer to the matrix. This is the reason for which it crucially depends on the heat transfer mechanism in the surrounding medium: while the Fourier law predicts a rapid decay, the BDE presents a much slower one. In fact, the Fourier theory is valid only when there are enough scattering events within the matrix. This assumption leads to an overestimation of the heat release from the particle when using Fourier's law, which is then to be evidently avoided in the cases of spatial nanoscale and/or subpicosecond heating processes. Therefore, the heat transfer from the particle is actually slower than estimated by the classical Fourier theory due to phonon rarefaction. The characteristic time for the slow relaxation is of course larger in the BDE case than in the Fourier one. In addition, as the energy injected by the pulse remains for a longer time in the particle, the lattice temperature predicted by the BDE is higher and its maximum is reached later than with Fourier's law (Fig. 2).

As the shell thickness is expected to play an important role in the heat transfer mechanism when its value is of the order of the phonon mean free path, let us investigate now its influence on the metal particle temperature relaxation. Figure 3 presents the electron temperature dynamics for different values of the shell thickness, $d_s$, from $0.1\Lambda_d = 0.54$ nm to $10\Lambda_d = 54$ nm. As can be seen on Fig. 3, $\Theta_e(t)$ is independent of $d_s$ during the short time regime (rapid relaxation). But after a few picoseconds, the profile of $\Theta_e(t)$ becomes thickness dependent: The thicker the shell, the larger the characteristic relaxation time.

This behavior may be explained as follows: the electron-phonon characteristic time being worth a few picoseconds, the energy transfer to the matrix is not effective during the short time regime. Consequently, $\Theta_e(t)$ is not sensitive to the morphology of the particle environment and the discussion given above regarding the characteristic time of the rapid relaxation remains valid. Now when the heat transfer to the matrix becomes effective, the shell thickness value can affect the particle temperature. Indeed, for a shell thickness inferior to or comparable with $\Lambda_d$, the heat transfer mainly follows a ballistic mechanism. When $d_s$ increases, the contribution of diffusive processes to the heat transfer becomes more important. This is shown in figure 4 where the time evolution of the diffusive phonon population at the particle surface is compared for different values of $d_s$. The diffusive phonon population around the particle is very low for a thin shell thickness, and its weight increases with thickness. It tends to a limit value when the shell thickness exceeds a few $\Lambda_d$. This behavior explains why the characteristic time in the domain of slow relaxation tends to a certain constant value (Fig. 3) corresponding to the "diffusive limit". For clarifying this point, Fig. 5 compares the time dependence of $\Theta_e$ obtained using the BDE and Fourier approaches for a shell thickness equivalent to $\Lambda_d$ and $10\Lambda_d$. As shown on this figure, the slow relaxation times predicted by these two approaches are quite different for $d_s = \Lambda_d$, while for $d_s = 10\Lambda_d$ they are similar. However, the particle temperature obtained using the BDE remains slightly higher than that provided by the Fourier law. This can be ascribed to the nonlocal heat transfer around the particle.[16] Note that the large thickness value corresponds to the case of a dilute nanocomposite medium, *i.e.* where the distance between two neighboring gold nanoparticles exceeds the path covered by the heat front within the time domain under consideration.

The relaxation rate also increases with the temperature gradient. The metal lattice temperature predicted by the BDE approach being higher than the one predicted by the Fourier model and the outer shell surface being thermalized, a larger temperature gradient appears around the metal particle. Consequently, the relaxation dynamics predicted by the BDE is faster (note the logarithmic vertical scale on Fig. 5). Let us also point out that the thicker the shell, the weaker the temperature gradient. This factor intervenes in both approaches (BDE and Fourier) and explains why the relaxation dynamics slows down as $d_s$ increases (Fig. 5).

It should also be mentioned that at every time $t$ before the heat front reaches the shell external surface, we may define an "effective thickness" corresponding to the path covered by the heat front from $t = 0$ to $t$.[26] Contrarily to the classical parabolic Fourier law, the existence of a zone in the medium not yet reached by the heat flux explains the similarity of the temperature profiles for $d_s = 5\Lambda_d$ and $10\Lambda_d$ on Fig. 3.

## b. Adiabatic boundary condition

Figures 6–8 correspond to Figures 2–4 when adiabatic boundary conditions are imposed on the shell outer surface. Under those conditions, the response of a close-packed core-shell particle system is predicted. Previous conditions of an imposed temperature at the outer boundary are more likely to describe dilute core-shell particle distributions in a highly conductive medium as a dielectric crystal. The differences between both sets of results are significant.

Figure 6 shows that BDE and Fourier responses have similar time dependence under adiabatic boundary conditions and when the shell thickness $d_s = \Lambda_d$. A quantitative disagreement of a few percents can still be observed. This difference is generated in a short time interval, apparently between 5 and 10 picoseconds, and it remains on longer times. In the BDE case, this effect is a signature of the ballistic flight of heat carriers or spatial non-locality. It might however not be easily detected through conventional pump-probe femtosecond laser experiments. Another striking point is that the electronic temperature profiles remain flat after 10 ps: The shell heat capacity being very small and the thermal energy being confined in the shell, the characteristic time for reaching the stationary regime approaches the phonon-electron relaxation time itself.

Figure 7 proves that a larger shell thickness leads to a stronger decrease of the electronic temperature because the total heat capacity of the shell is augmented. When $d_s = 10\Lambda_d$ the difference between BDE and Fourier predictions remains the same as for $d_s = \Lambda_d$ in terms of absolute values but the relative difference this time reaches 10-20%. The boundary conditions do not affect significantly this last response because the ballistic contribution cancels out when heat carriers undergo several collisions or cross several mean free paths.

Finally, Figure 8 reveals more than one order of magnitude difference between the time responses of the electronic temperature when shell thickness varies. The reason for this behaviour is once again that thermal energy is confined in the shell. The apparition of the plateau only starts when the shell is isothermal. The larger are the shell thickness and heat capacity, the later the plateau appears. In the case of imposed temperature conditions, stationary regime only establishes when heat flux cancels, *i.e.* when temperature on the shell external surface reaches the imposed temperature.

## IV. CONCLUSION

According to previous works, the heat transfer from the particle to the matrix can deviate from Fourier's law prediction depending on particle size: the smaller the particle, the stronger the deviation[16]. Consequently, the influence of the non-Fourier heat transport on the temperature dynamics linked with particle size may be envisaged for very small particles. However, we have not observed any non-Fourier deviation for particles with radii larger than 2 nm ($\approx 0.34 \Lambda_d$) in the case of the alumina matrix. In this case indeed, such influence should be revealed only for sub-nanometer particles. Such small particles are not relevant for optical investigations because of their very weak surface plasmon resonance due to electron confinement effects.

It should also be reminded that the non-Fourier heat transport due to the shell and possibly to the particle is related to the nonlocal transport mechanisms linked with the nanoscale structure of the material. Another non-Fourier heat transport mechanism is related to the short heating time. This effect becomes significant when the latter is inferior to the phonon relaxation time τ in the shell medium. Here the heating time is governed by the electron-phonon coupling time of a few picoseconds and neither by the pulse temporal width of 110 fs nor by τ. Such influence would however manifest itself in the slow relaxation dynamics if the phonon relaxation time τ in the shell was larger.

We would finally like to underline two points regarding the model. The first one concerns the difference between the boundary conditions in the BDE description and in the Fourier one. In fact, the values of the temperatures used in the boundary conditions for the BDE are those of the emitted phonons (non-reflecting condition), while in the Fourier approach they are those of equilibrium phonons. This causes a temperature jump at the interface in the BDE (or BTE) case.[18,19,26] This temperature jump is thus linked with the ballistic origin of the phonons at the metal-dielectric interface and depends on the shell thickness: The thinner the shell, the stronger the temperature jump.[26]

The second point is related to the limit of the ballistic-diffusive approach. As the diffusive contribution in the BDE is only an approximation to the heat carrier scattering processes, this model is less accurate when the diffusive component is predominant and particularly in the steady state.[18,19]

In summary, we have modeled the physical mechanisms governing the electron and lattice temperature dynamics in a gold nanoparticle core embedded in a dielectric shell under a subpicosecond laser pulse, considering two different external boundary conditions. This model is based on the usual electron-phonon coupling in the particle and the ballistic-diffusive approximation in the dielectric medium. It is well suited to such situation where the Fourier theory is disqualified to describe the heat propagation correctly at the space and time scales involved. For a thin dielectric shell, in the case of imposed temperature conditions at the outer boundary, the prediction of the Fourier law for the slow relaxation characteristic time deviates from that predicted using the BDE. This non-Fourier effect related to spatial non-locality is a valuable mean to determine the thickness of the shell or its phonon mean free path.

As the shell thickness increases, the thermal behavior tends to a "diffusive limit" for which the slow relaxation time is comparable with that obtained using the Fourier approach for heat conduction. However, the lattice temperature determined through the BDE is higher than the one obtained from Fourier's law. This result induces a significant impact on the interpretation of pump-probe laser experiments involving nanoparticle based composites.

Non-Fourier heat transport affecting the dynamics of the particle temperature may also occur (*i*) for very small particles with sizes much smaller than $\Lambda_d$ and (*ii*) when the particle heating rate is comparable or inferior to the phonon relaxation time in the matrix, $\tau$. Confirming these results by ultrafast pump-probe experiments now represents an interesting issue.

**Acknowledgements**. The financial supports of the Agence Nationale de la Recherche (program ANR/PNANO 2006, project EThNA) and the region Ile-de-France, under the project SESAME E-1751, are gratefully acknowledged.

**APPENDIX**

Results are presented in terms of the following non-dimensional parameters:

- electron non-dimensional temperature: $\Theta_e(t) = \dfrac{T_e(t) - T_0}{T_0}$,

- lattice non-dimensional temperature: $\Theta_l(t) = \dfrac{T_l(t) - T_0}{T_0}$,

395     - non- dimensional heat fluxes: $q_b^*(r,t) = \dfrac{q_b(r,t) - q_{0b}}{C_d |\mathbf{v}| T_0}$ and $q_m^*(r,t) = \dfrac{q_m(r,t) - q_{0m}}{C_d |\mathbf{v}| T_0}$,

- non- dimensional internal energies: $\Theta_b(r,t) = \dfrac{u_b(r,t) - u_{0b}}{C_d T_0}$ and

$\Theta_m(r,t) = \dfrac{u_m(r,t) - u_{0m}}{C_d T_0}$ .

**END APPENDIX**

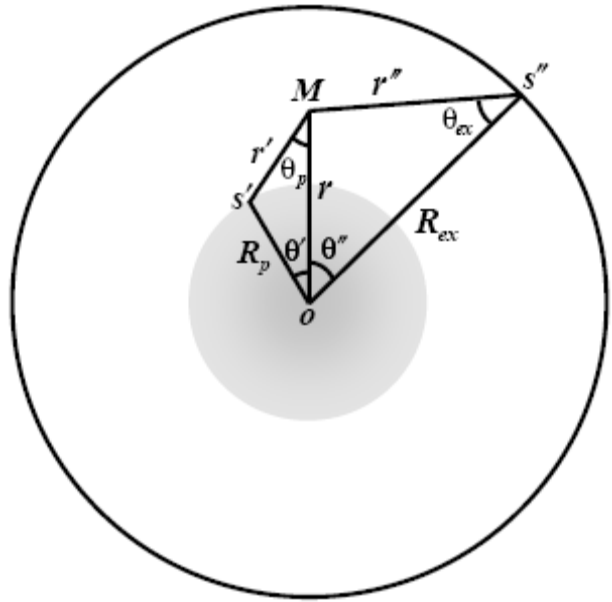

Figure 1: Schematic diagram of a metal particle, with radius $R_p$, surrounded by a dielectric shell with external radius $R_{ex}$. $r$ is the distance of a point $M$ in the shell from the particle centre. $r'$ and $r''$ represent the distances between $M$ and a surface element of the particle ($s'$) or that of the shell external surface ($s''$), respectively. $\theta_p$ and $\theta_{ex}$ are the polar angles of these two surface elements relative to (*MO*).

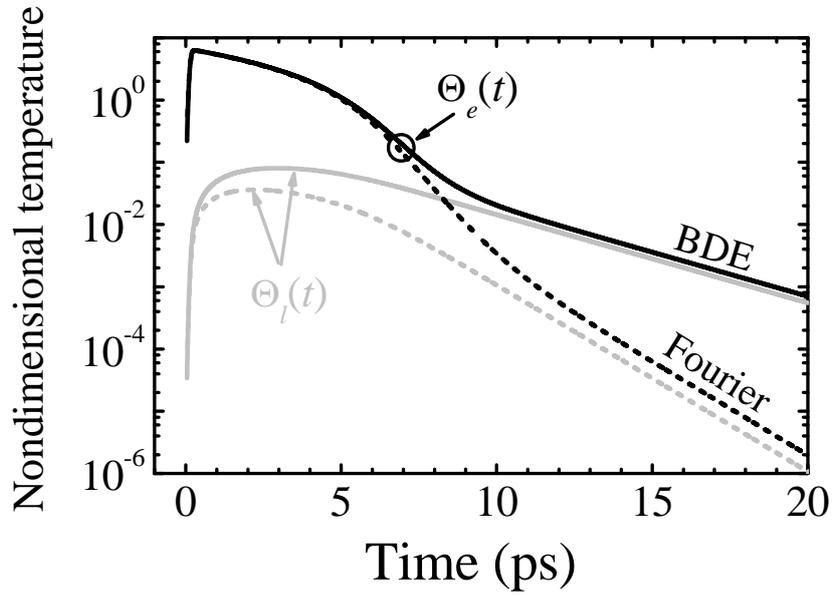

410

Figure 2: Non-dimensional electron and lattice temperature dynamics [$\Theta_e(t)$ and $\Theta_l(t)$, respectively] of a gold nanoparticle surrounded by an alumina shell under an ultrashort laser pulse excitation. The data are obtained using the ballistic-diffusive approximation (solid lines) and Fourier's law (dash lines) within the thermalizing boundary condition (imposed temperature at the shell outer surface). The nanoparticle radius is 10 nm and the shell thickness is equal to the phonon mean free path in the dielectric $\Lambda_d \approx 5.4$ nm. The laser pulse duration is 110 fs.

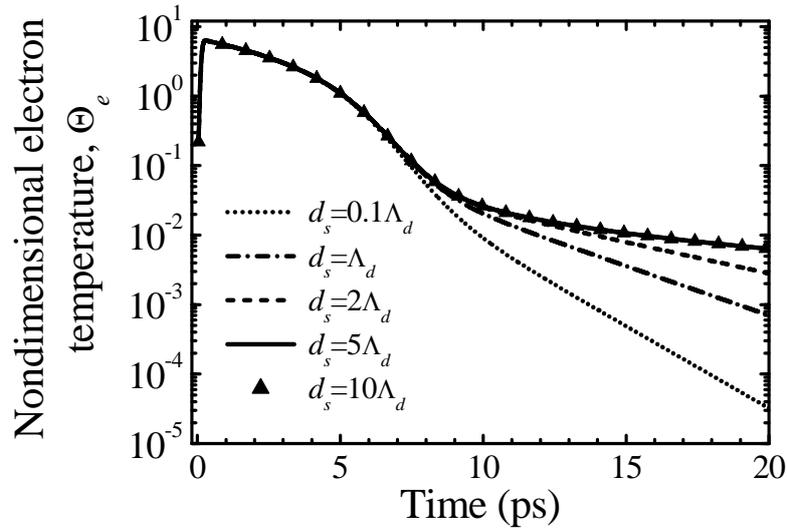

Figure 3: Non-dimensional electron temperature dynamics of a gold nanoparticle (10 nm radius) in an alumina shell, as calculated by solving the BDE with the thermalizing boundary condition. Shell thickness $d_s$ varies from $0.1\Lambda_d \simeq 0.54$ nm to $10\Lambda_d \simeq 54$ nm.

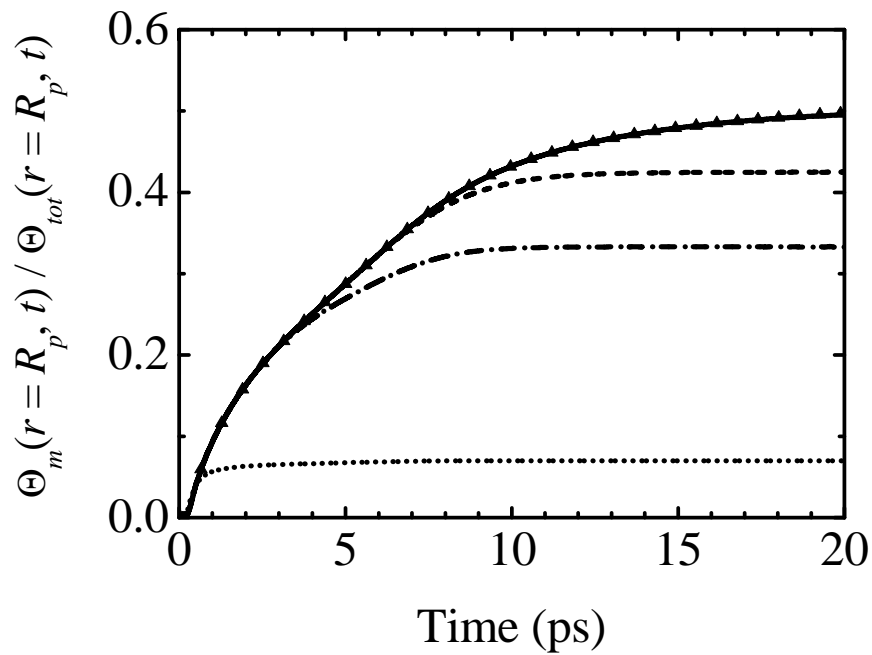

Figure 4: Relative diffusive phonon population at the particle surface for different shell thicknesses (see caption of Fig. 3).

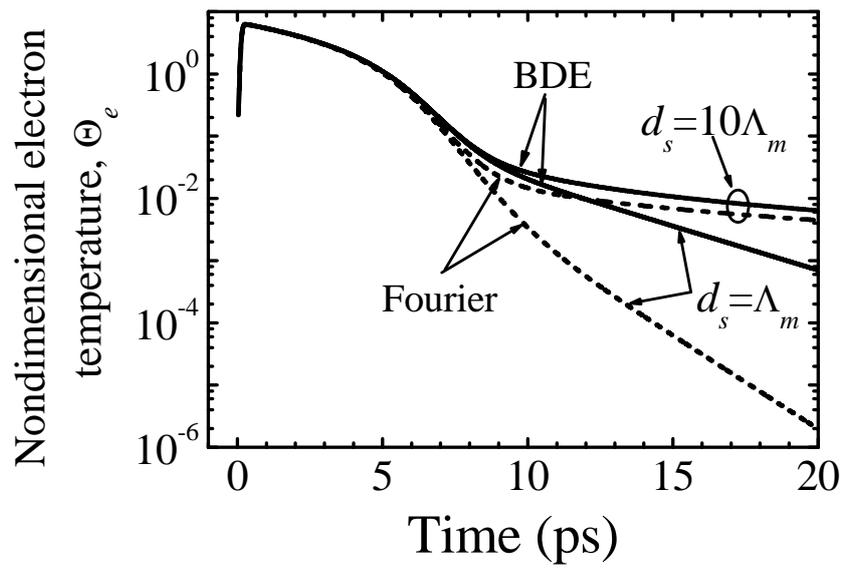

430  Figure 5: Non-dimensional electron temperature dynamics of a gold nanoparticle (10 nm radius) in an alumina shell with two different thickness values, obtained using the BDE (solid lines) and Fourier's law (dash lines) with the thermalizing boundary condition.

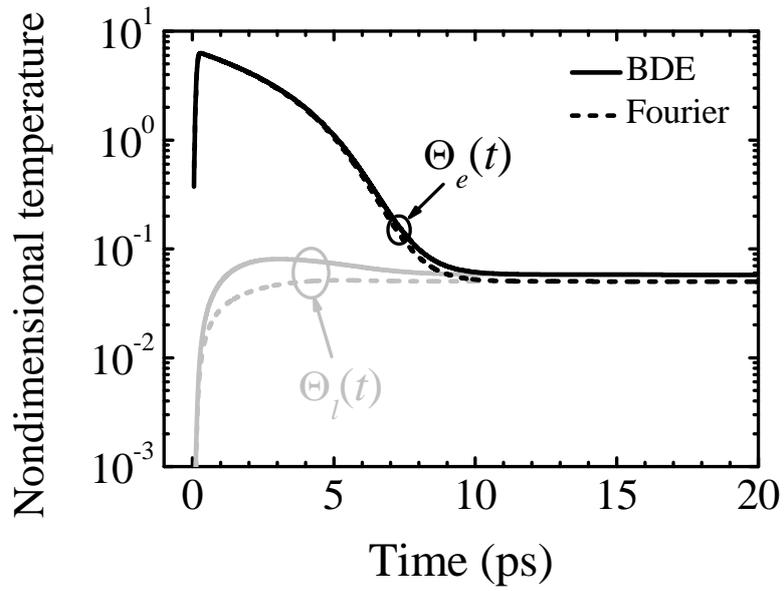

Figure 6: Same as Fig. 2 with adiabatic condition at the shell outer boundary.

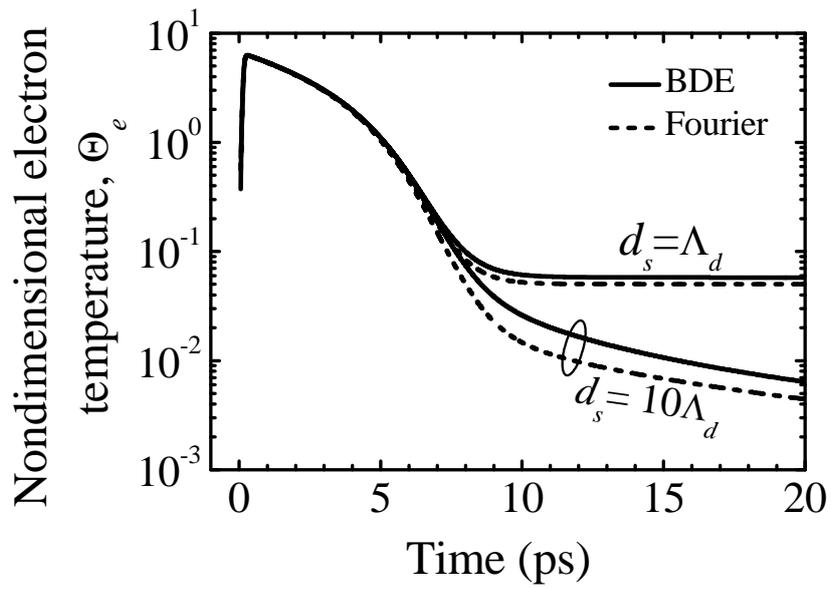

Figure 7: Same as Fig. 4 with adiabatic condition.

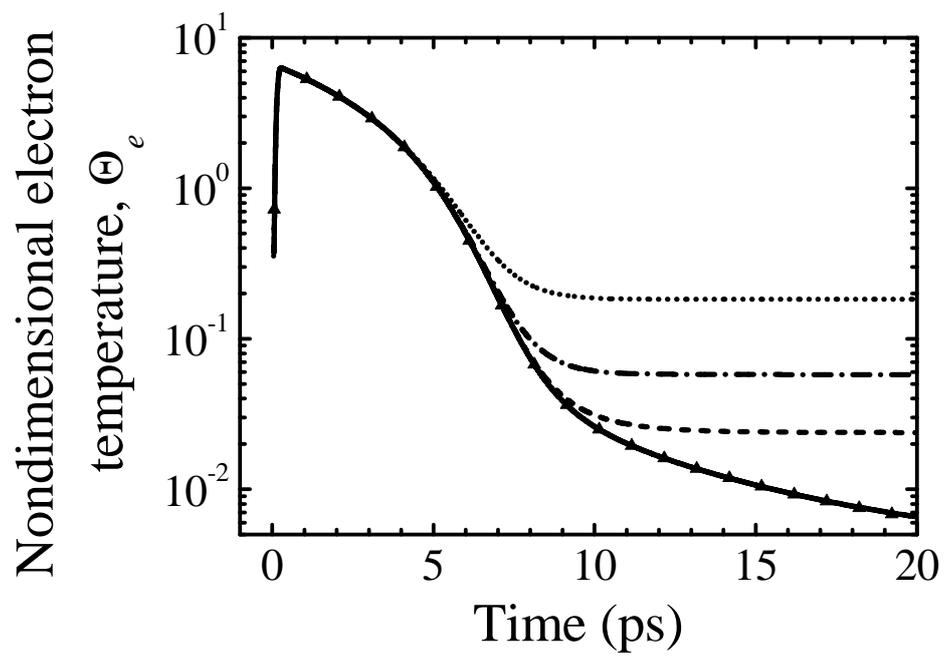

  Figure 8: Same as Fig. 3 with adiabatic condition.